\documentstyle[aps]{revtex}
\begin{document}
\draft
\title{
FROM CONSTITUENT QUARK TO HADRON STRUCTURE IN THE NEXT-to-LEADING ORDER:
NUCLEON AND PION
}

\author{Firooz Arash$^{(a)}$\footnote{e-mail: arash@vax.ipm.ac.ir} and
Ali Naghi Khorramian$^{(b)}$}
\address{
$^{(a)}$
Institute for Studies in Theoretical Physics and Mathematics
 P.O.Box 19395-5531, Tehran, Iran \\
$^{(b)}$
Physics Department, Amir Kabir University of Technology,
Hafez Avenue, Tehran, Iran \\
}
\date{\today}
\maketitle
\begin{abstract}
We calculate the partonic structure of constituent quark in the Next-to-Leading
Order for the first time. The structure of any hadron can be obtained thereafter
using a convolution method. Such a procedure is used to generate the structure
function of proton and pion in NLO. It is shown that while the constituent
quark structure is generated purely perturbatively and accounts to most part of the
hadronic structure, there is a few percent contributions coming from the
nonperturbative sector in the hadronic structure. This contribution plays the
key role in explaining the $SU(2)$ symmetry breaking of the nucleon sea and the
observed violation of Gottfried sum rule. These effects are calculated. Excellent
agreement with data in a wide range of $x=[10^{-6}, 1]$ and $Q^{2}=[0.5, 5000]$
$GeV^{2}$ is reached for proton structure function. We have also calculated
Pion structure and compared it with the existing data. Again nice agreement is
achieved. {\bf PACS Numbers:  13.60 Hb, 12.39.-x, 13.88 +e, 12.20.Fv}
\end{abstract}
\section{INTRODUCTION}
Our knowledge of hadronic structure is based on the hadronic spectroscopy and
the Deep Inelastic Scattering (DIS) data. In the former picture quarks are massive
particles and their bound states describe the static properties of hadrons;
while the interpretation of DIS data relies upon the quarks of QCD Lagrangian
with very small mass. The hadronic structure in this picture is intimately
connected with the presence of a large number of partons(quarks and gluons).
The two types of quarks not only differ in mass, but also on other important
properties. for example, the color charge of quark field in QCD Lagrangian is
ill-defined and is not gauge invariant, reflecting the color of gluons in an
interacting theory. On the other hand, color associated with a Constituent
Quark (CQ) is a well defined entity. It has been recently shown that \cite{1}\cite{2}
one can perturbatively dress a QCD Lagrangian field to all orders and construct
a CQ in conformity with the color confinement. From this point of view a CQ is
defined as a quasi-particle emerging from the dressing of valence quark with
gluons and $q-\bar{q}$ pairs in QCD.\\
Of course, the concept of CQ as an intermediate step between the quarks of QCD
Lagrangian and hadrons is not new. In fact, in the context of $SU(6) X O(3)$
long before Altarelli and Cabibo \cite{3} have used them. R.C. Hwa in his
elaborated work termed them as {\it{valon}}, extended it and showed its
application to many physical processes \cite{4}. In Ref.[2] it is suggested
that the concept of dressed quark and gluon might be useful in the area of
jet physics and heavy quark effective theory. Despite the ever presence of CQ
no one has calculated its content and partonic structure without resorting
to hadronic data and the process of deconvolution. The purpose of this paper
is threefold: (a) we will evaluate the structure of a CQ in the
Next-to-Leading Order (NLO) in QCD. (b) we will verify its conformity with the
structure function data of nucleon and pion for which there are ample data
available. (c) In the process, however, we will notice that additional refinements
are needed to account for the violation of Gottfried Sum Rule (GSR) and the
effect of binding of CQ's to form a physical hadron. \\
\section{FORMALISM}
By definition, a CQ is a universal building block for every hadron, that is,
its structure is common to all hadrons and generated perturbatively. Once its
structure is evaluated, in principle it would permit to calculate the structure
of any hadron. In doing so we will follow the philosophy that in a DIS experiment
at high enough $Q^{2}$ it is the structure of a CQ which is being probed and at
sufficiently low value of $Q^{2}$ this structure cannot be resolved thus, a CQ
behaves as valence quark and hadron is viewed as the bound state of its CQ's.
Under these criteria partons of DIS experiments are components of CQ and at
high $Q^{2}$ one can write for a U-type CQ its structure as follows:
\begin{equation}
F_{2}^{U}(z,Q^2)=\frac{4}{9}z(G_{\frac{u}{U}}+G_{\frac{\bar{u}}{U}})+ \frac{1}
{9}z(G_{\frac{d}{U}}+G_{\frac{\bar{d}}{U}}+G_{\frac{s}{U}}+G_{\frac{\bar{s}}{U}})+...
\end{equation}
where all the functions on the right-hand side are the probability functions for
quarks having momentum fraction $z$ of a U-type constituent quark at $Q^{2}$.
Similar expression can be written for a D-type CQ. 
Defining the singlet (S) and nonsinglet (NS) CQ distribution functions as:
\begin{equation}
G^{S}=\sum^{f}_{i=1}(G_{\frac{q_{i}}{CQ}} + G_{\frac{\bar{q}_{i}}{CQ}}) =G_{f} +(2f-1)G_{uf}
\end{equation}
\begin{equation}
G^{NS}=\sum^{f}_{i=1}(G_{\frac{q_{i}}{CQ}} -G_{\frac{\bar{q}_{i}}{CQ}}) =G_{f} - G_{uf}
\end{equation}
where $G_{f}$ is the favored distribution describing the structure function of
a quark within a CQ of the same flavor while unfavored distribution $G_{uf}$
describes the structure function of any quark of different flavor within the CQ.
$f$ is the number of active flavors. In terms of singlet and nonsinglet
distributions they are as follows:
\begin{equation}
G_{f}=\frac{1}{2f}(G^{S} + (2f-1) G^{NS})
\end{equation}
\begin{equation}
G_{uf}=\frac{1}{2f}(G^{S}-G^{NS})
\end{equation}
Having expressed all the structure functions of CQ's in terms of $G_{f}$ and
$G_{uf}$ we now go to the moment space and define the moments of these
distributions as:
\begin{equation}
M(n,Q^{2})=\int_{0}^{1} x^{n-2} F(x,Q^{2}) dx
\end{equation}
\begin{equation}
M_{i}(n,Q^{2})=\int_{0}^{1} x^{n-1} G_{i}(n,Q^{2})dx
\end{equation}
the subscription $i$ stands for S, NS. Using charge symmetry, in the following
we will refer to CQ distribution only in proton. \\
In the NLO approximation the dependence of the running coupling constant, $\alpha$
on $Q^{2}$ is given by:
\begin{equation}
\alpha(Q^{2})=\frac{4\pi}{\beta_{0}ln(\frac{Q^{2}}{\Lambda^{2}})}
\big(1-\frac{\beta_{1}ln ln(\frac{Q^{2}}{\Lambda^{2}})}
{\beta_{0}^{2}ln(\frac{Q^{2}}{\Lambda^{2}})}\big)
\end{equation}
with $\beta_{0}=\frac{1}{3} (33-2f)$ and $\beta_{1}=102-\frac{38f}{3}$. The
moments of NS and S in the NLO are:
\begin{equation}
M^{NS}(n,Q^{2})= [1+\frac{\alpha(Q^{2})-\alpha(Q_{0}^{2})}{4\pi}\big(
\frac{\gamma^{(1)N}_{NS}}{2\beta_{0}}-\frac{\beta_{1}\gamma_{qq}^{(0)N}}{2\beta_{0}^{2}}
\big)](\frac{\alpha_{s}(Q^{2})}{\alpha_{s}(Q_{0}^{2})})^{\gamma_{qq}^{(0)N/2\beta_{0}}}
\end{equation}
\begin{eqnarray}
M^{S}(n,Q^{2})=\{(\frac{\alpha_{s}(Q^{2})}{\alpha_{s}(Q_{0}^{2})})^{\frac{\lambda_{-}^{N}}{2\beta_{0}}}[
p_{-}^{N}-\frac{1}{2\beta_{0}}\frac{\alpha_{s}(Q^{2}_{0})-\alpha_{s}(Q^{2})}{4\pi}p_{-}^{N}
\gamma^{N}p_{-}^{N}-\big(\frac{\alpha_{s}(Q^{2}_{0})}{4\pi}-\frac{\alpha_{s}(Q^{2})}{4\pi}
)^{\frac{\lambda_{+}^{N}-\lambda_{-}^{n}}{2\beta_{0}}}\big) \nonumber \\
\frac{p_{-}^{N}\gamma^{N}p_{-}^{N}}{2\beta_{0}^{2}+\lambda^{N}_{+}-\lambda^{N}_{-}}]+(+\longleftrightarrow -)\}
\end{eqnarray}
where $\gamma^{N}=\gamma^{(1)N}-\frac{\beta_{1}}{\beta_{0}}\gamma^{(0)N}$ and
$\gamma^{(0)N}$ and $\gamma^{(1)N}$ are anomalous dimension metrices\cite{5} and
$\lambda_{\pm}^{N}$ denote the eigenvalues of one-loop anomalous dimension matrix
$\gamma^{(0)N}$.
\begin{equation}
\lambda_{\pm}^{N}=\frac{1}{2}[\gamma_{qq}^{(0)N}+\gamma_{gg}^{(0)N}\pm \sqrt{
(\gamma_{gg}^{(0)N}-\gamma_{qq}^{(0)N})^{2}+4\gamma_{qg}^{(0)N}\gamma_{gq}^{(0)N}}]
\end{equation}
and $p_{\pm}^{N}$ is given by:
\begin{equation}
p_{\pm}^{N}=\pm(\gamma^{(0)N}-\gamma_{\pm}^{N})/(\lambda_{+}^{N}-\lambda_{-}^{N})
\end{equation}
The quantities $d_{\pm}^{(0)}$, the leading order anomalous
dimentions, given in Ref.[4], are related to $\lambda_{\pm}$ in our notations.
$t$ is the evolution parameter defined as:
\begin{equation}
t={\it{ln}}\frac{\it{ln}\frac{Q^2}{\Lambda^2}}{\it{ln}\frac{Q_{0}^{2}}{\Lambda^2}}
\end{equation}
The coefficients $\gamma_{kl}^{(1,0)N}$ and can be found in \cite{6}. \\
We have taken our initial scale $Q^{2}_{0}=0.283$ $GeV^{2}$ and $\Lambda=0.22$ GeV.
It seems that evolution of parton distributions from such a low value of
$Q^{2}_{0}$ is not justified theoretically.
The above value of $Q_{0}$ corresponds to a distance of 0.36 {\it{fm}} which is
roughly equal to or slightly less than the radius of a CQ. It may be objected
that such a distances are probably too large for a meaningful pure perturbative
treatment. We note that $F_{2}^{CQ}(z,Q^{2})$ has the property that it becomes
$\delta(z-1)$ as $Q^{2}$ is extrapolated to $Q_{0}^{2}$, which is beyond the
region of validity. This mathematical boundary condition signifies that the
internal structure of a CQ cannot be resolved at $Q_{0}$ in the NLO approximation.
Consequently, when this property is applied to Eq.(19) bellow, the structure function
of the nucleon becomes directly related to $xG_{\frac{CQ}{P}}(x)$ at those
values of $Q_{0}$, that is, $Q_{0}$ is the leading order effective value at
which the hadron can be considered as consisting only of three (two) CQ's, for
baryons (mesons). In fact our results are only meaningful for $Q_{0}^{2}\ge 0.4$
$GeV^{2}$. As it is stated above, the moments of the CQ structure function,
$F_{2}^{CQ}(z,Q^{2})$ are expressed completely in terms of the evolution
parameter, $t$, of Eq. (13). From the theoretical standpoint, both $\Lambda$
and $Q_{0}$ depend on the order of the moments $n$. In this work we have
assumed that they are independent of $n$, hence introducing some degrees of
approximation to the $Q^{2}$ evolution of the valence and sea quarks. However,
on one hand there are other contributions like target-mass effects, which add
uncertainties to the theoretical predictions of perturbative QCD, while on the
other hand since we are dealing with the CQ, there is no experimental data
to invalidate an $n$ independent $\Lambda$ assumption.
The moments of valence and sea quarks in a CQ are:
\begin{equation}
M_{\frac{valence}{CQ}}=M^{NS}(n,Q^{2})
\end{equation}
\begin{equation}
M_{\frac{sea}{CQ}}=\frac{1}{2f}(M^{S}-M^{NS})
\end{equation}
where $M^{S,NS}$ are given above. Evaluating $M_{\frac{valence}{CQ}}$ and
$M_{\frac{sea}{CQ}}$ at any $Q^{2}$ or $t$ is now straight forward. Using Inverse
Mellin Transform techniques, following forms for the valence and sea quark
distributions inside a CQ is obtained in the NLO:
\begin{equation}
zq_{\frac{val.}{CQ}}(z,Q^{2})=a z^{b}(1-z)^{c}
\end{equation}
\begin{equation}
zq_{\frac{sea}{CQ}}(z,Q^{2}) = \alpha z^{\beta}(1-z)^{\gamma}[1+\eta z +\xi z^{0.5}]
\end{equation}
The parameters $a$, $b$, $c$ , $\alpha$, etc. are functions of $Q^{2}$ through
the evolution parameter $t$. The same form as in Eq.(17) is obtained for Gluon
distribution in a CQ but only with different parameters. Functional form of them
is a polynomial of order three in $t$ and are given in the appendix. We notice
that the following sum rule reflecting the fact that each CQ contains only one
valence quark is satisfied for all values of $Q^{2}$:
\begin{equation}
\int^{1}_{0} q_{\frac{val.}{CQ}}(z,Q^{2}) dz = 1.
\end{equation}
Substituting these results in Eq.(1) completes the evaluation of a constituent
quark structure function in NLO. In Figure (1) various parton distributions inside a
CQ is plotted.
\section{HADRONIC STRUCTURE}
In previous section we calculated the NLO structure of a CQ. In this section
we will use the convolution theorem to calculate the structure function of
proton, $F_{2}^{p}(x,Q^{2})$, and that of a pion. Let us denote the structure
function of a CQ by $F^{CQ}_{2}(z,Q^{2})$ and the probability of finding a CQ
carrying momentum fraction $y$ of the hadron by $G_{\frac{CQ}{h}}(y)$. The
corresponding structure function of the hadron, using the convolution theorem
is as follows:
\begin{equation}
F_{2}^{h}(x,Q^2)=\sum_{CQ}\int_{x}^{1}\frac{dy}{y} G_{\frac{CQ}{h}}(y)F^{CQ}_{2}(\frac{x}{y},Q^2)
\end{equation}
where summation runs over the number of CQ's in a particular hadron. Also notice
that $G_{\frac{CQ}{h}}(y)$ is independent of the nature of the probe and its
$Q^{2}$ value. $G_{\frac{CQ}{h}}(y)$ in effect, describes the wave function of
hadron in CQ representation containing all the complications due to confinement.
From the theoretical point of view this function cannot be evaluated accurately.
To facilitate phenomenological analysis, following Ref.[4] we assume a simple
form for the exclusive CQ distribution in proton and pion as follows:
\begin{equation}
G_{UUD/p}(y_{1},y_{2},y_{3})=l(y_{1}y_{2})^{m}y_{3}^{n}\delta(y_{1}+y_{2}+y_{3}-1)
\end{equation}
\begin{equation}
G_{\bar{U}D/\pi^{-}}(y_{1},y_{2})=qy_{1}^{\mu}y_{2}^{\nu}\delta(y_{1}+y_{2}-1)
\end{equation}
Integrating over unwanted momenta, we can arrive at inclusive distribution of
individual CQ:
\begin{equation} 
G_{U/p}(y)=\frac{1}{B(a+1,b+a+2)}y^{a}(1-y)^b 
\end{equation}
\begin{equation}
G_{D/p}(y)=\frac{1}{B(b+1,2a+2)}y^{b}(1-y)^{2a+1}
\end{equation}
\begin{equation}
G_{\bar{U}/\pi^{-}}(y)=\frac{1}{B(\mu +1, \nu +1)}y^{\mu}(1-y)^{\nu}
\end{equation}
similarly expression for $G_{D/\pi^{-}}$ with the interchange of
$\mu\leftrightarrow \nu$. In the above equations $B(i,j)$ is Euler Beta function
and its arguments are fixed using the sum rule:
\begin{equation}
\int^{1}_{0}G_{\frac{CQ}{h}}(y) dy=1
\end{equation}
where $CQ=U, D, \bar{U}$ and $h=p,\pi^{-}$. Numerical values are: $\mu=0.01$,
$\nu =0.06$, $a=0.65$ and $b=0.35$. In Figure (2) the CQ distributions in proton
and $\pi^{-}$ are shown. We stress that CQ distributions in hadrons are independent
of $Q^{2}$ and the nature of probe being used. Now it is possible to determine
various parton distributions in a hadron. For proton we can write:
\begin{equation}
q_{val./p}(x,Q^{2})=2\int^{1}_{x}\frac{dy}{y} G_{U/p}(y) q_{val./U}(x,Q^{2})
+\int^{1}_{x}\frac{dy}{y} G_{D/p}(y) q_{val./D}(x,Q^{2}) =u_{val./p}(x, Q^{2}) +d_{val./p}(x, Q^{2})
\end{equation}
\begin{equation}
q_{sea/p}(x,Q^{2})=2\int^{1}_{x}\frac{dy}{y} G_{U/p}(y) q_{sea/U}(x,Q^{2}) +
\int^{1}_{x}\frac{dy}{y} G_{D/p}(y) q_{sea/D}(x,Q^{2})
\end{equation}
The above equation represents the contribution of constituent quarks to the nucleon sea.
Comparing with data on proton structure functions shows that the results fall
short of representing the experimental data by a 3-5 percent. This is due in
our opinion, to the fact that CQ is not free in a hadron but they interact with
each other in forming the bound states. That means, there are some soft gluons
in the nucleon besides the CQ's. In the process of formation of bound state,
a CQ emits gluon which in turn decays into $\bar{q}-q$ pairs which gives a
residual component to the partons in a hadron. In our picture there is no room
in the CQ structure for breaking the $SU(2)$ symmetry of the sea but after
creation of $\bar{q}-q$ pair from the emitted gluon , these quarks can recombine
with CQ to fluctuate into meson-nucleon state which breaks the symmetry of the
nucleon see. In what follows we will compute this component, its contribution
to the nucleon structure function and violation of Gottfried sum rule following
the preswcription used in \cite{7} \cite{8}. In order to
distinguish these partons from those confined inside the CQ, we will term them as
{\it{inherent partons}}. Although this component is intimately related to the
bound state problem, and hence it has a non-perturbative origin, for not so
small values of $Q^{2}$ we will calculate it perturbatively for the process of
$CQ\rightarrow CQ+{\it{gluon}}\rightarrow \bar{q}-q$ at an initial value of
$Q^{2}=0.65 GeV^{2}$ where $\alpha_{s}$ is still small enough. The corresponding
splitting functions are as follows:
\begin{equation}
P_{gq}(z)=\frac{4}{3}\frac{1+(1-z)^{2}}{z}
\end{equation}
\begin{equation}
P_{qg}(z)=\frac{1}{2}(z^{2}+(1+z)^{2})
\end{equation}
For the joint probability distribution of the process at hand, we get:
\begin{equation}
q_{inh.}(x,Q^2)=\bar{q}_{inh}(x,Q^2)=N\frac{\alpha^{2}_{s}}{(2\pi)^2}\int_{x}^{1}\frac{dy}{y}P_{qg}(\frac{x}{y})\int_{y}^{1}\frac{dz}{z}P_{gq}(\frac{y}{z})G_{CQ}(z)
\end{equation}
The splitting functions and the $q_{inh}(x,Q^2)=\bar{q}_{inh}(x,Q^2)$, above are
that of the leading order rather than NLO. We do not expect it should make much
of a difference since, its contribution to the whole structure function is only
a few percent as can be seen in Figure (3). In the above equation $N$ is a factor
depending on $Q^{2}$ and $G_{CQ}$ is the constituent quark distribution in the
proton given previously. The same process, however, also can be a source of
$SU(2)$ symmetry breaking of nucleon sea and resulting in $u_{sea}\neq d_{sea}$, and
hence the violation of Gottfried sum rule (GSR). There are several explanations
forthis observation such as flavor asymmetry of the nucleon sea \cite{9} \cite{10}, isospin
symmetry breaking between proton and neutron, Pauli blocking, {\it{etc}}. One
of these explanations fits well within our model. It was proposed  by Eichten, Inchliffe
and Quigg \cite{11} that valence quark fluctuates into quark and a pion. In other
words a nucleon can fluctuate into a meson-nucleon state. This idea is appealing
in our model and can be calculated rather easily. In our model after a pair of
{\it{inherent}} $q-\bar{q}$ created a $\bar{u}$ can couple to a D-type CQ to form
an intermediate $\pi^{-}=D\bar{u}$ while the $u$ quark combines with the other
two U-type CQ's to form a $\Delta^{++}$. This is the lowest $u\bar{u}$ fluctuation.
Similarly a $d\bar{d}$ can fluctuate into the $\pi^{+}n$ state. Since
$\Delta^{++}$ state is more massive than $n$ state, then the probability of $d\bar{d}$
fluctuation will dominate over $u\bar{u}$ fluctuation which naturally leads to
an excess of $d\bar{d}$ pairs over $u\bar{u}$ in the proton sea. This process is
depicted in Figure (4). Probability of formation of a meson-barion state can be
written as in Ref. [8]:
\begin{equation}
P_{MB}(x)=\int^{1}_{0}\frac{dy}{y}\int^{1}_{0}\frac{dz}{z}F(y,z)R(y,z;x)
\end{equation}
where $F(y,z)$ is the joint probability of finding a CQ with momentum fraction
$y$ and an inherent quark or anti-quark of momentum fraction $z$ in the proton.
$R(y,z;x)$ is the probability of recombining a CQ of momentum $y$ with an
inherent quark of momentum $z$ to form a meson of momentum fraction $x$ in the proton.
The evaluation of both of these probability functions are discussed in \cite{12}
for a more general case and an earlier, but pioneering, version also proposed in
\cite{13}. In the present case these functions are much simpler. Guided by works
done in Ref. [8,12,13] we can write :
\begin{equation}
F(y,z)=\Omega yG_{\frac{CQ}{p}}(y)z\bar{q}_{inh.}(z)(1-y-z)^{\delta}
\end{equation}
\begin{equation}
R(y,z;x)=\rho y^{a}z^{b}\delta(y+z-1)
\end{equation}
Here we take $a=b=1$ reflecting that two CQ's in meson almost equally share
its momentum. The exponent $\delta$ is fixed for the $n$ and $\Delta^{++}$ states
using the data from E866 experiment and the mass ratio of $\Delta$ to $n$. They
turn out to be approximately 18 and 13 respectively. $\Omega$ and $\rho$ are the
normalization constants also fixed by data. It is now possible to evaluate
$\bar{u}_{M}$ and $\bar{d}_{M}$ quarks associated with the formation of meson states:
\begin{equation}
\bar{d}_{M}(x,Q^{2}) =\int^{1}_{x}\frac{dy}{y}[P_{\pi n}(y) +
\frac{1}{6}P_{\pi \Delta ^{++}}(y)]d_{\pi}(\frac{x}{y},Q^{2})
\end{equation}
\begin{equation}
\bar{u}_{M}(x,Q^{2})=\frac{1}{2}\int^{1}_{x}\frac{dy}{y}
P_{\pi \Delta^{++}}(y)u_{\pi}(\frac{x}{y},Q^{2})
\end{equation}
where $u_{\pi}$ and $d_{\pi}$ are the valence quark probability
densities in the pion at scale $Q^{2}_{0}$. The coefficients $\frac{1}{2}$
and $\frac{1}{6}$ are due to Isospin consideration. Using Eqs. (16, 17, 24) we can
calculate various parton distributions in a the pion. Those pertinent to Eqs. (34, 35)
are:
\begin{equation}
\bar{u}_{val.}^{\pi^{-}}(x, Q^{2})=\int^{1}_{x}G_{\bar{U}/\pi^{-}}(y)
\bar{u}_{val./\bar{U}}(\frac{x}{y},Q^{2})\frac{dy}{y}
\end{equation}
\begin{equation}
d_{val.}^{\pi^{-}}(x, Q^{2})=\int^{1}_{x}G_{D/\pi^{-}}(y)
d_{val./D}(\frac{x}{y},Q^{2})\frac{dy}{y}
\end{equation}
\begin{equation}
\bar{u}_{val./\bar{U}}=u_{val./U}
\end{equation}
There are some data on the valence structure function of $\pi^{-}$ \cite{14}.
Defining valence structure function of $\pi^{-}$ as:
\begin{equation}
F^{\pi^{-}}_{val.}=x\bar{u}^{\pi^{-}}_{val.} =xd^{\pi^{-}}_{val.}
\end{equation}
we present the results of our calculation for $F^{\pi^{-}}_{val.}$ in Figure (5)
along with the experimental data at $Q^{2}$ around 6 $Gev^{2}$. Returning to the
$\bar{d}$ excess over $\bar{u}$ in proton we can write all the contributions as:
\begin {equation}
(\frac{\bar{d}}{\bar{u}})_{proton}=\frac{\bar{d}_{M}+\bar{d}_{inh.+CQ}}
{\bar{u}_{M}+\bar{u}_{inh.+CQ}}
\end{equation}
NuSea collaboration at FermiLab E866 experiment has published their results \cite{15}
for integral of $\bar{d}-\bar{u}$ and $\frac{\bar{d}}{\bar{u}}$ at $Q=7.35$ $GeV$.
With the procedure described, we have calculated these values at the same $Q$
and for the range of $x$ as experimentally measured: $x=[0.02, 0.35]$. The results of
the model is:
\begin{equation}
\int^{0.345}_{0.02}dx (\bar{d}-\bar{u})=0.085
\end{equation}
to be compared with the experimental value of $0.068\pm 0.0106$. We get for the
entire range in $x$: $\int^{1}_{0}dx (\bar{d}-\bar{u})=0.103$ while the
experimentally extrapolated value is $0.1 \pm 0.018$ which are in excellent agreement
with our calculations. This gives a value for the Gottfried sum rule of:
\begin{equation}
S_{G}=\int_{0}^{1}[F_{2}^{p}(x)-F_{2}^{n}(x)]\frac{dx}{x}=
\frac{1}{3}-\frac{2}{3}\int_{0}^{1}dx [\bar{d}(x)-\bar{u}(x)]=0.264.
\end{equation}
at $Q=7.35$ $GeV$. The NMC result \cite{14} is $S_{G}=0.235 \pm 0.026$ which is at
much lower value of $Q^{2}=4$  $GeV^{2}$. In Figure (6), $\bar{d}(x)-\bar{u}(x)$
and  $\frac{\bar{d}}{\bar{u}}$ in proton are shown as a function of $x$ at
$Q=7.35$ $GeV$ along with the measured results. \\
We are now in a position to present the results for the proton structure
function, $F_{2}^{p}$. In Eqs.(16,17) we presented the form of parton
distributions in a CQ. Using those relations with the numerical values given
in appendix then from Eq.(1) the structure function of CQ is obtained. In Eqs.
(22, 23) the shape of CQ distributions in proton is given. With the help of
Eq. (19) now all the ingredients are in place to calculate $F_{2}^{p}$. In
Figure (7) the results are shown at many values of $Q^{2}$. As it is evident
they fall a few percent short of representing the data. However, as mentioned
earlier, there is an additional contribution from the {\it{inherent}} partons to $F^{p}_{2}$
which is calculated in Eq. (30). Adding this component represents the data
rather well and can be seen from Figure (7). The data points are from \cite{17}.
For the purpose of comparing
our results with other calculations, we have also included in Figure (7), the
GRV's NLO results \cite{18} as well as the prediction of CTEQ4M \cite{19}.
Notice that we have taken the number of active flavors
to be three for $Q^{2}\leq 5 GeV^{2}$ and four flavors elsewhere. In Figure (8)
the gluon distribution predicted by the model is presented along with those
from Ref.[18, 19].
\section{summary and conclusion}
In this paper we have used the notion of constituent quark as a well defined
entity being common to all hadrons. Its structure can be calculated in QCD
perturbatively to all orders. A CQ receives its own structure by dressing a
valence quark with gluon and $q-\bar{q}$ pairs in QCD. We have calculated its
structure function in the Next-to-Leading order for the first time. Considering
a hadron as the bound states of these CQs we have used the convolution theorem
to extract the hadronic structure functions for proton and pion. Besides the
CQ structure contribution to the hadrons, there is also a nonperturbative
component while contributing only a few percent to the overall structure of
hadrons becomes crucial in explaining the violation of Gottfried sum rule and
the excess of $\bar{d}$ over $\bar{u}$ in the nucleon sea. A mechanism is
devised for this purpose and necessary calculations are outlined. We have
presented the results and compared them with all available relevant data and
with the work of others. We found that our results are in good agreement with the
data.
\section{APPENDIX}
In this appendix we will give the functional form of parameters of Eqs. (16, 17)
in terms of the evolution parameter, $t$. This will completely determines
partonic structure of CQ and their evolution. The results are valid for three
and four flavors, although the flavor number is not explicitly present but
they have entered in through the calculation of moments. As we explained in the
text, we have taken the number of flavors to be three for $Q^{2} \leq 5 GeV^{2}$
and four for higher $Q^{2}$ values. \\
\\
I) Valence quark in CQ (Eq. 16): \\ 
\\
$a= -0.1512 +1.785 t -1.145 t^{2} +0.2168 t^{3}$  \\
$b=1.460 -1.137 t +0.471 t^{2} -0.089 t^{3}$   \\
$c=-1.031 +1.037 t -0.023 t^{2} +0.0075 t^{3}$ \\
\\
II) Sea quark in CQ (Eq. 17):  \\
\\
$\alpha=0.070 - 0.213 t + 0.247 t^{2} - 0.080 t^{3}$ \\
$\beta=0.336 - 1.703 t +1.495 t^{2} - 0.455 t^{3}$ \\
$\gamma=-20.526 +57.495 t -46.892 t^{2} + 12.057 t^{3}$ \\
$\eta =3.187 - 9.141 t +10.000 t^{2} -3.306 t^{3}$ \\
$\xi=-7.914 +19.177 t - 18.023 t^{2} + 5.279 t^{3}$ \\
$N=1.023 +0.124 t -2.306 t^{2} +1.965 t^{3}$ \\
\\
III) Gluon in CQ (Eq. 17) \\
\\
$\alpha=0.826 - 1.643 t + 1.856 t^{2} - 0.564 t^{3}$ \\
$\beta=0.328-1.363 t + 0.950 t^{2} -0.242 t^{3}$ \\
$\gamma=-0.482 + 1.528 t -0.223 t^{2} -0.023 t^{3}$ \\
$\eta=0.480 -3.386 t + 4.616 t^{2} - 1.441 t^{3}$  \\
$\xi=-2.375 + 6.873 t -7.458 t^{2} +2.161 t^{3}$ \\
$N=2.247-6.903 t + 6.879 t^{2} -1.876 t^{3}$ \\

\newpage 
\section{Figure Caption}
Figure-1. Moments of partons in a CQ at $Q^{2}=20$ $GeV^{2}$ as a function of $z$. \\
Figure-2. Parton distributions in proton and $\pi^{-}$ at $Q^{2}=20$ $GeV^{2}$ as
a function of $x$. \\
Figure-3. Contribution of {\it{inherent}} component to a sea parton distribution.
The dashed-dotted line is that of CQ and the solid line represents the sum of
the two components. \\
Figure-4. Processes responsible to $SU(2)$ symmetry breaking in the nucleon sea
and violation of Gottfried sum rule. \\
Figure-5. Pion valence structure function as a functuon of $x$ at
$Q^{2}=5.5$ $(Gev/c)^2$. Solid curve is the result of the model calculations
and the data points are from Ref.[12]. \\
Figure-6. The ratio $\frac{\bar{d}}{\bar{u}}$ and the difference $\bar{d}-\bar{u}$
as a function of $X$. The solid line in the model calculation and the dotted
line is the prediction of CTEQ4M. Data are from Ref. [13]. \\
Figure-7. Proton structure function $F_{2}^{p}$ as a function of $x$ calculated
using the model and compared with the data from Ref. [15] for different $Q^{2}$ values.
The {\it{thin line}} is the prediction of GRV Ref.[16] and the dashed line is
that of CTEQ4M Ref.[17]. \\
Figure-8. The gluon distribution in proton as a function of $x$ at $Q^{2}=20$ $GeV^{2}$.
We have also shown the prediction of GRV (dashed-dotted line) and CTEQ4M (dashed line).
The data points are from H1 collaboration.
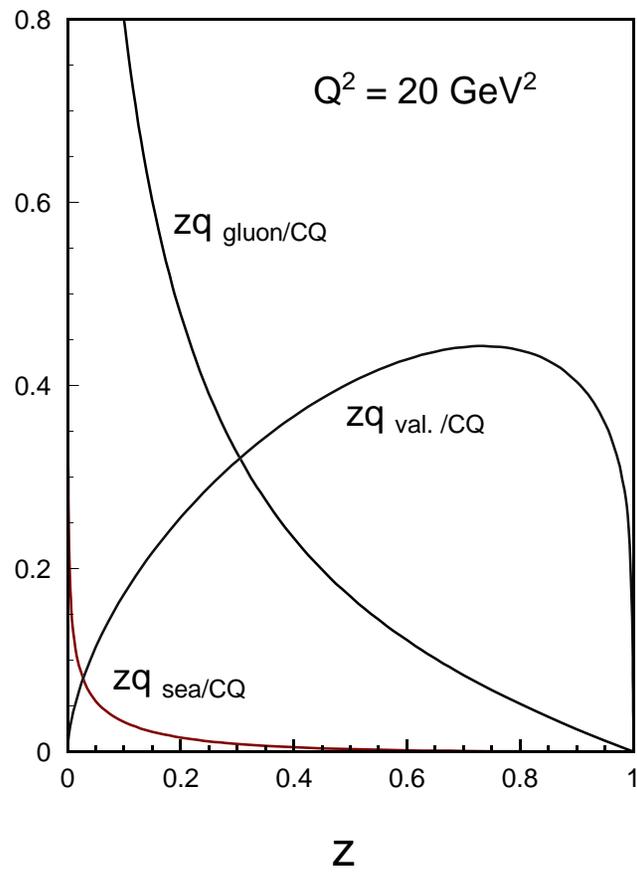
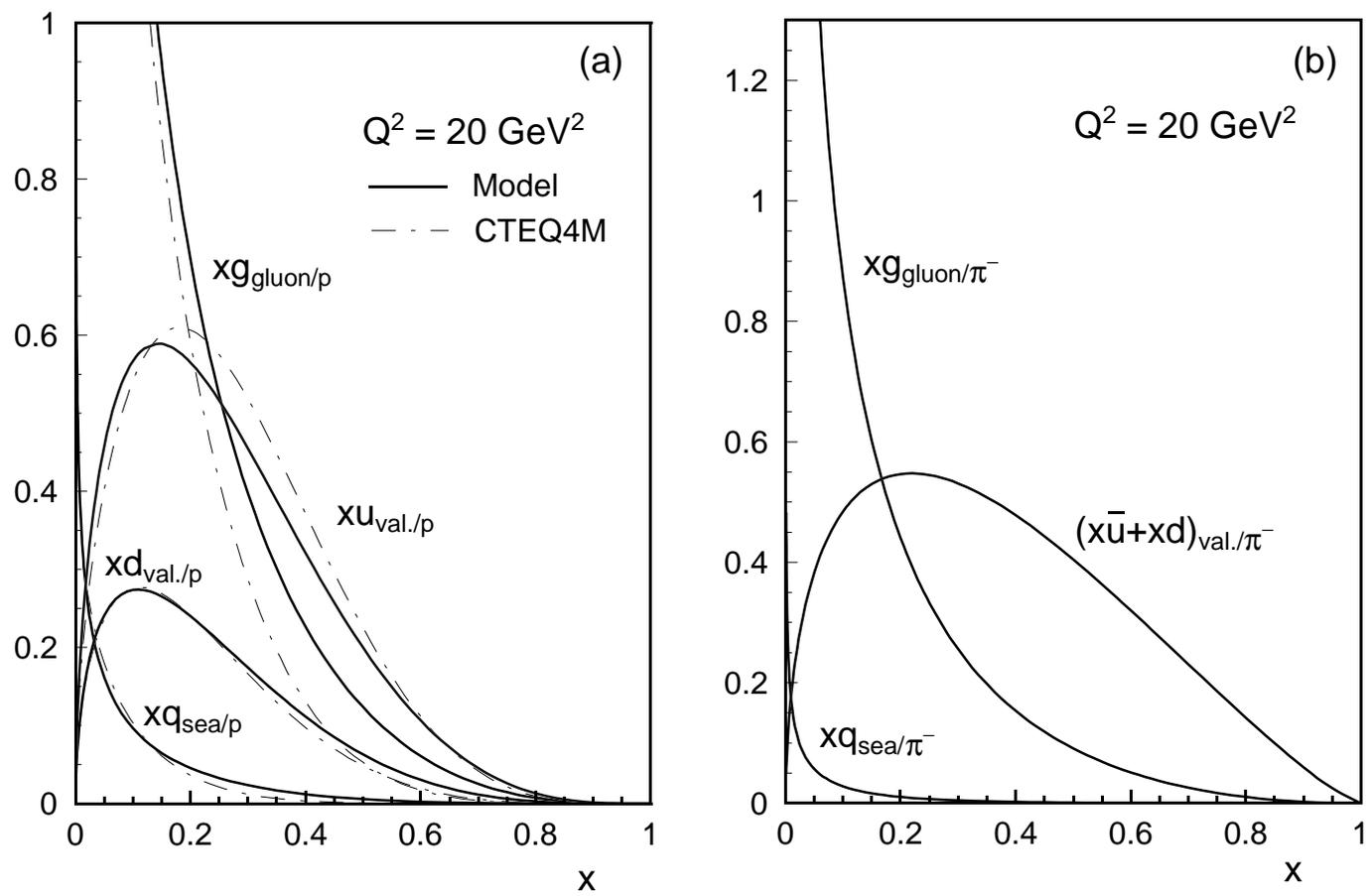
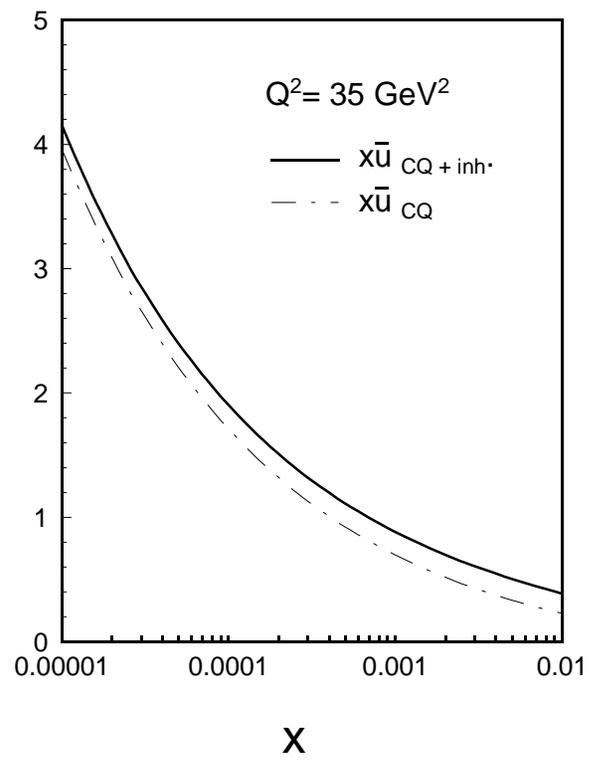
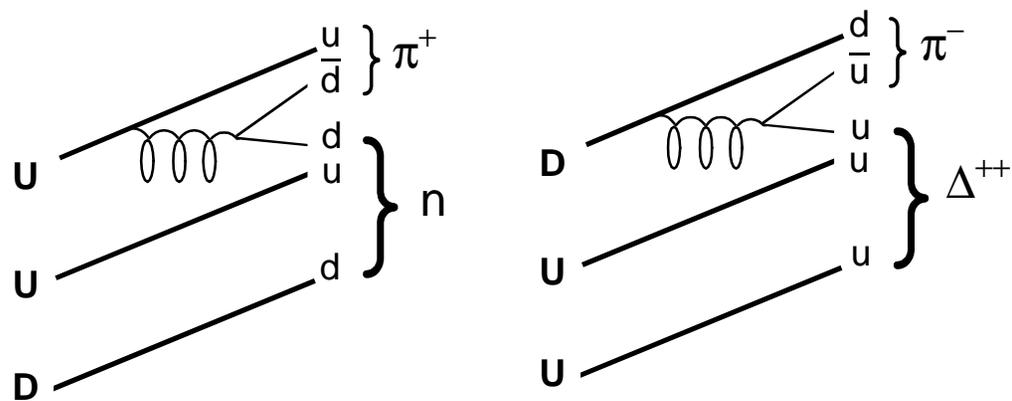
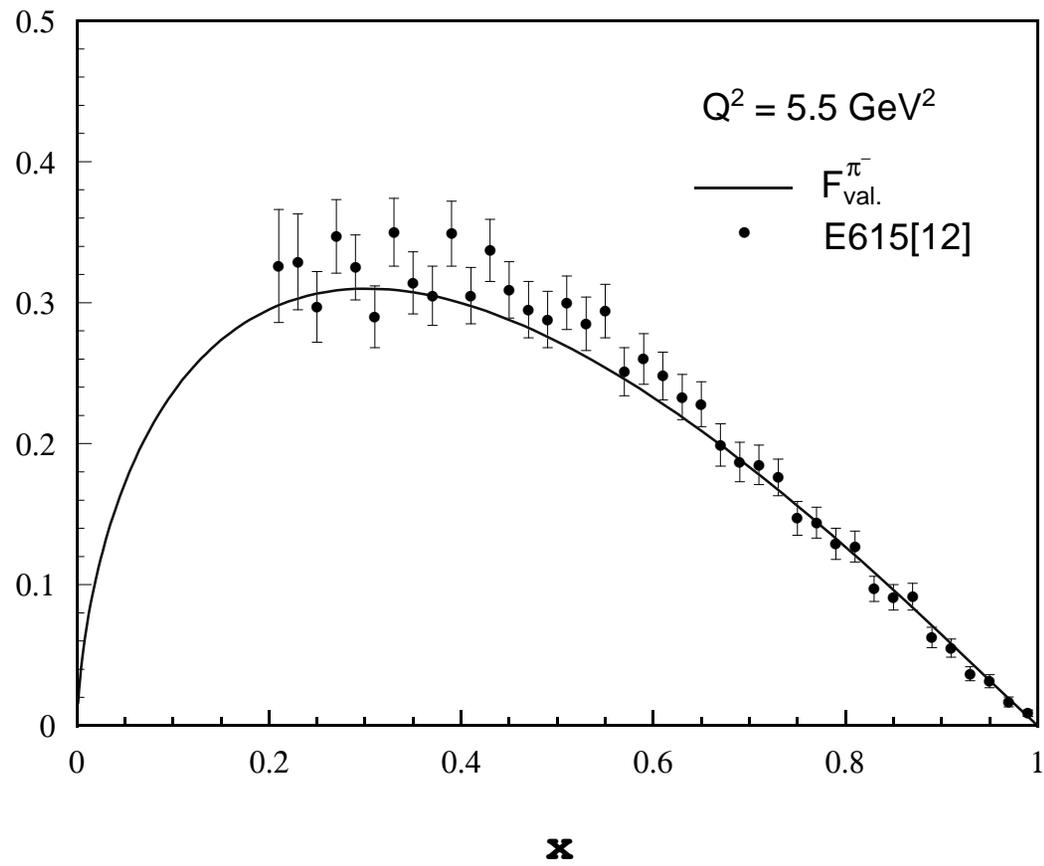
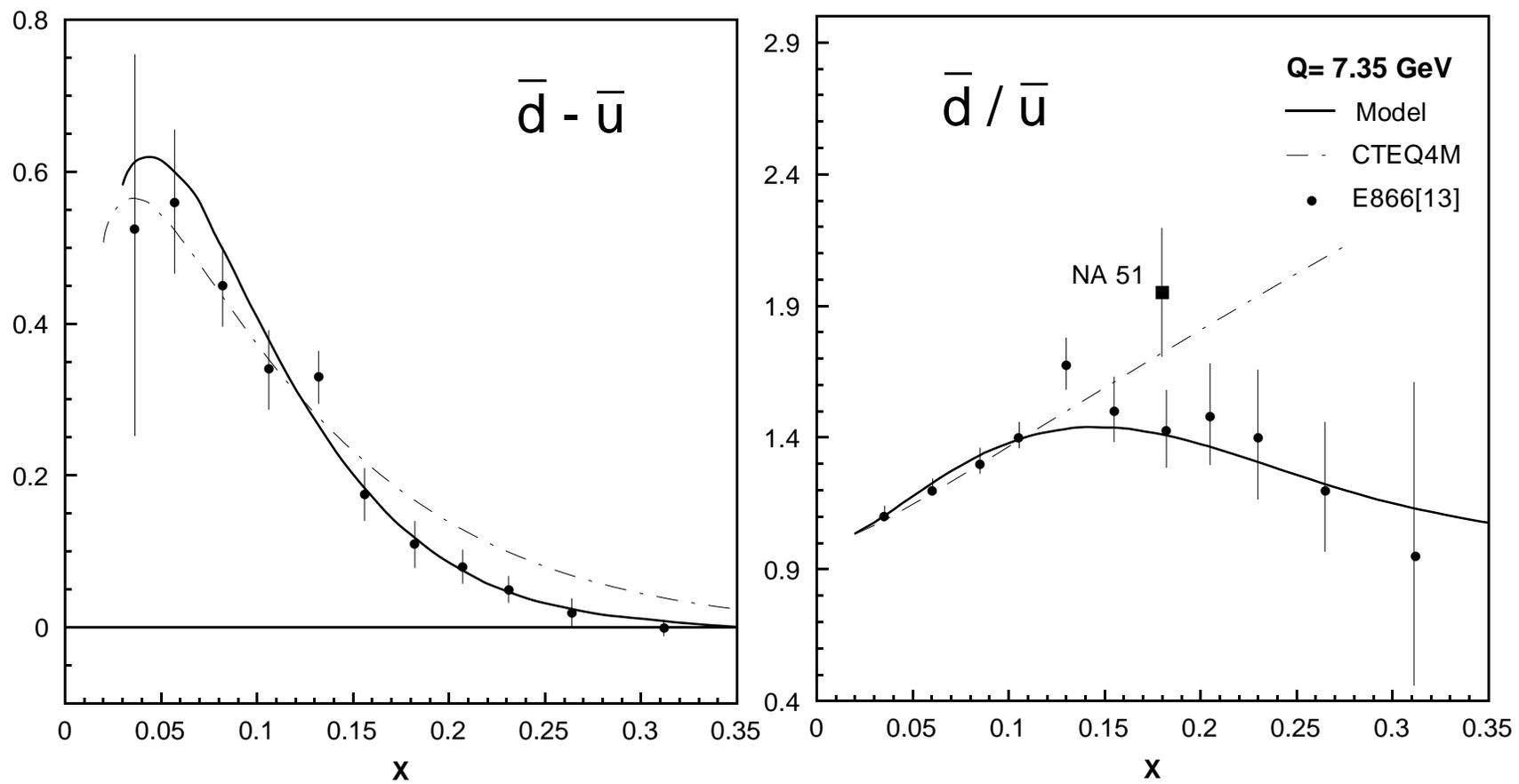
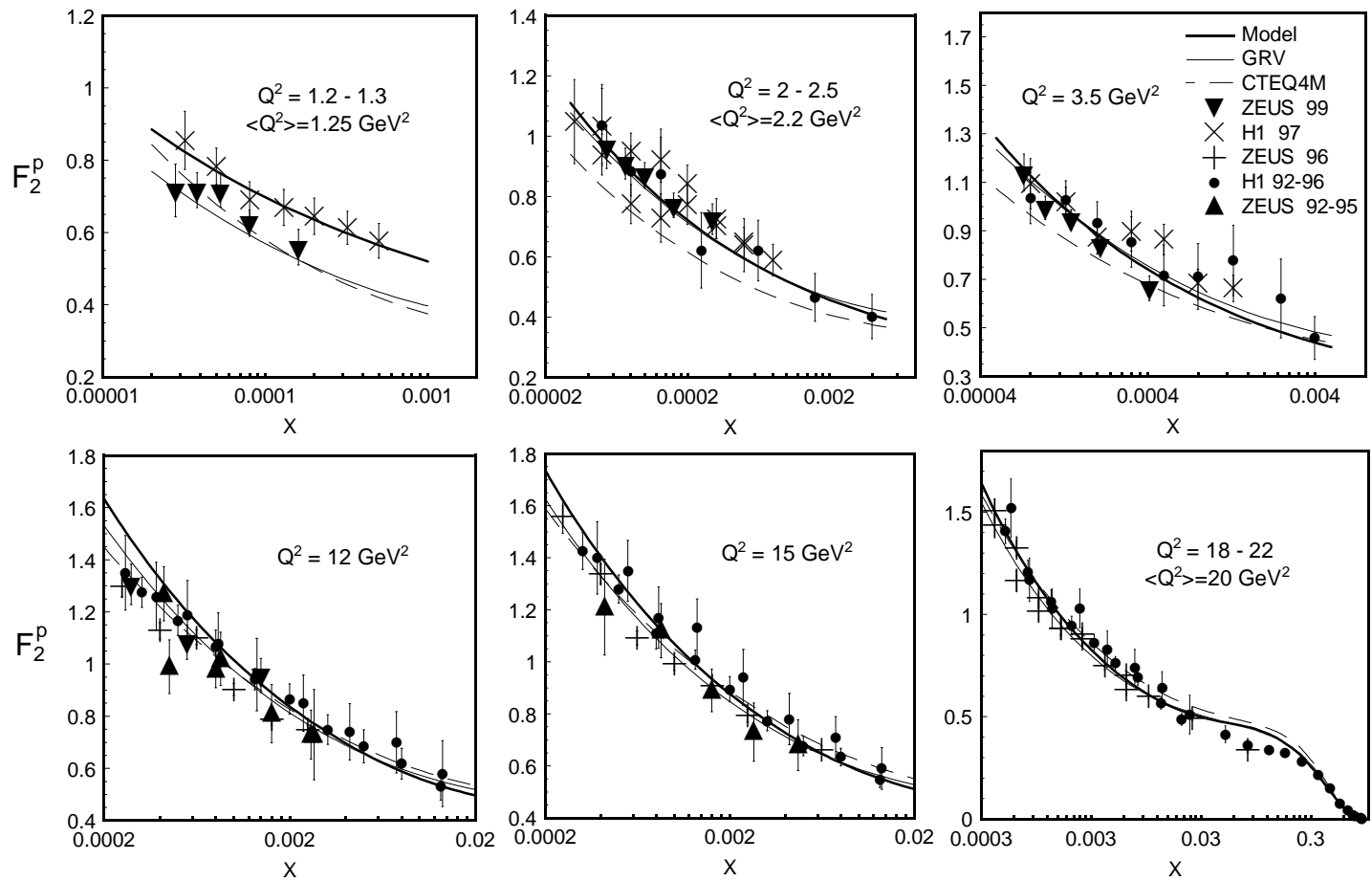
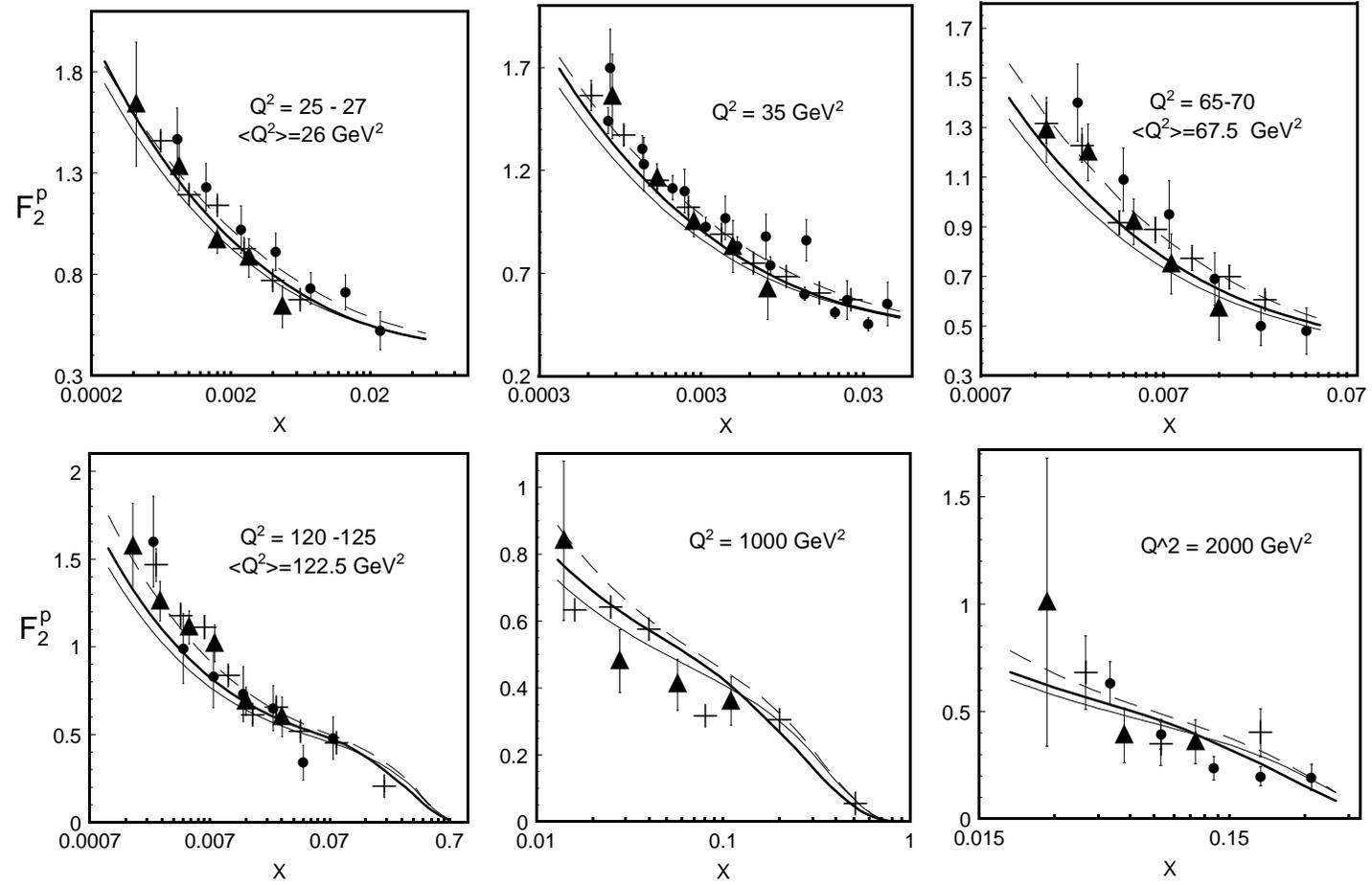
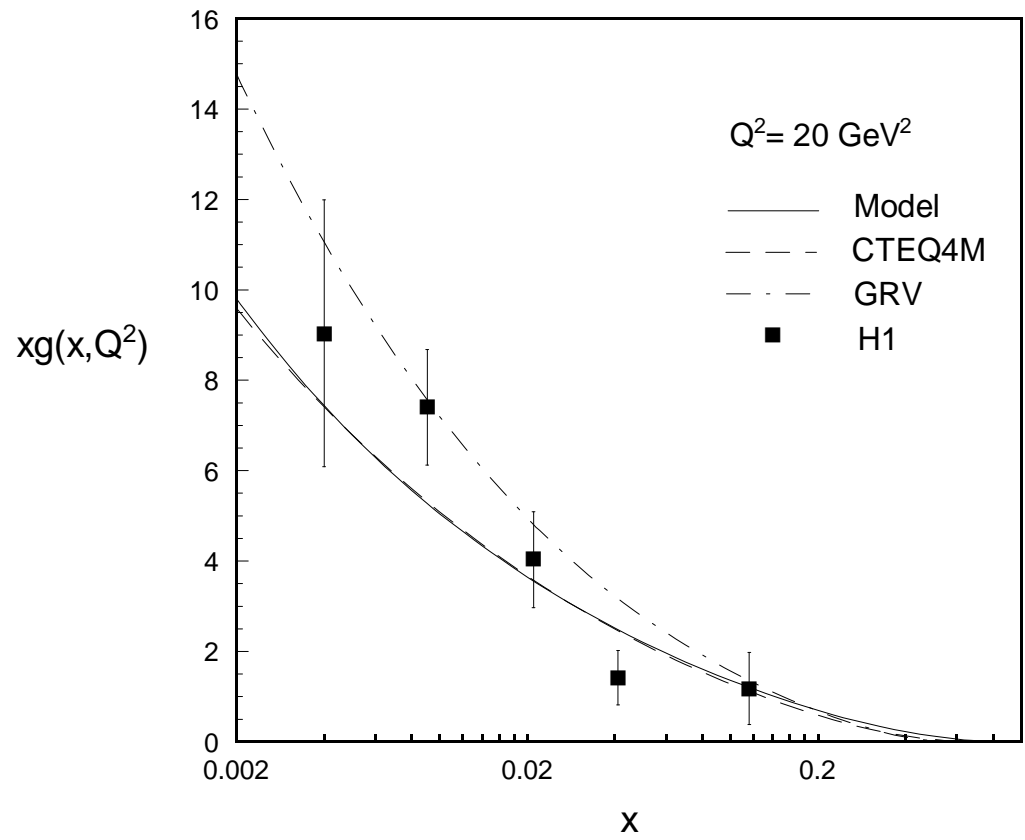
\end{document}